\documentstyle[12pt]{article}
\def\lsim{\mathrel{\rlap{\raise 2.5pt \hbox{$<$}}\lower 2.5pt}}
\def\gsim{\mathrel{\rlap{\raise 2.5pt \hbox{$>$}}\lower 2.5pt}}
\begin{document}
%\begin{titlepage}
\thispagestyle{empty}
\begin{small}
\begin{flushright}
BUTP-95/38, KL-TH~95/25, hep-ph/9511415\\
\end{flushright}
\end{small}
\vspace{-3mm}
\begin{center}
{\bf{\Large Particle Spectrum in the Non-Minimal Supersymmetric \\
Standard Model with $\tan\beta\simeq m_t/m_b$}}
\vskip 0.5cm
B. Ananthanarayan\\[-2mm]
\begin{small}
Institut de physique th\'eorique, Universit\'e de Lausanne,\\ [-2mm]
CH 1015, Lausanne, Switzerland.\\[-2mm]

Institut f\"ur Theoretische Physik, Universit\"at Bern,\\[-2mm]
CH 3012, Bern, Switzerland.$^*$\\[-2mm]
\vskip 0.2cm
\end{small}
P. N. Pandita\\[-2mm]
\begin{small}
Universit\"at Kaiserslautern, Fachbereich Physik,\\[-2mm]
Erwin--Schr\"odinger--Strasse, D-67663 Kaiserslautern, Germany.\\[-2mm]

Department of Physics, North Eastern Hill University,\\[-2mm]
Laitumkhrah, Shillong 793 003, India.$^{**}$\\[-2mm]
%\vskip 2.5cm
\end{small}
\end{center}
%\vskip 2in
%\newpage
\begin{abstract}
We present a detailed  discussion of the particle spectrum of the
Non-Minimal Supersymmetric
Standard Model (NMSSM), containing two Higgs
doublets and a singlet, in the limit $\tan\beta \simeq m_t/m_b$.
This is
compared with  the corresponding particle spectrum of the
Minimal Supersymmetric
Standard Model (MSSM).  In this limit the singlet vacuum
expectation value is forced to be large, of the order of $10$ TeV, and
the singlet decouples from the lightest scalar Higgs boson and
the neutralinos.  With the exception of the lightest Higgs boson, the
particle spectrum in the model turns out to be heavy.  The radiatively
corrected lightest
Higgs boson mass is in the neighbourhood of $\sim 130$ GeV.
\end{abstract}
\begin{small}
{\it PACS number(s): 12.60.Jv, 12.10.Kt, 14.80.Ly}\\[-2mm]
\noindent{\underline{\hspace{11.6cm}}}\\[-2mm]
* Present address\\[-2mm]
** Permanent address
\end{small}

%\end{titlepage}

\newpage
%\dspace
\noindent{\bf 1. Introduction.}

\bigskip

The Non-Minimal Supersymmetric Standard Model (NMSSM) [1], where the
particle content of the Minimal Supersymmetric Standard Model (MSSM) [2]
is extended by the addition of a gauge singlet chiral superfield $S$, and
one in which dimensionful couplings are eliminated through the introduction
of a discrete $Z_3$ symmetry, offers an interesting and viable alternative
to the MSSM.  It has a significantly richer phenomenology and a typically
larger parameter space.  Furthermore, with the prospect of LHC running
in the not so distant future, it is important to consider variations
of the MSSM in order to test the stability of its predictions so that
search strategies may be appropriately extended or modified in order to
discover low energy supersymmetry.

The predictiveness of the MSSM is vastly enhanced when the crucial
 parameter $\tan\beta\equiv
v_2/v_1$, the ratio of the vacuum expectation values of
the two Higgs doublets $H_2$ and $H_1$, required to give
masses to the up-type and the down-type (and charged
leptons) quarks, respectively, is constrained via the requirement
that the b-quark mass come out in the right range [3,4] in
supersymmetric unification [5]
(our normalization is
$v\equiv\sqrt{v_1^2+v_2^2}=174 GeV$, and the mass of the Z boson
is defined such that
$m_Z^2=\frac{1}{2}(g^2+g'^2)v^2$, where $g$ and $g'$ are
the gauge couplings of $SU(2)$ and $U(1)$, respectively).
One particularly predictive framework
is based on the assumption that the heaviest generation
fermions lie in a unique {\bf 16}-dimensional representation
of the unifying gauge group $SO(10)$ with the Higgs doublets
in a {\bf 10}-dimensional representation of the group[3].  This
implies that the top-quark, b-quark and $\tau$ lepton
Yukawa interactions arise from a $h. {\bf 16}.{\bf 16}.{\bf 10}$ term in
the superpotential at the unification scale $M_X$ determined
from gauge coupling unification.
The coupled system of differential equations for
the gauge couplings and Yukawa couplings are then
evolved down to present energies from $M_X$,
and $\tan\beta$ is determined from the accurately
measured value of $m_\tau=1.78\ GeV$.  When $h(=h_t=
h_b=h_\tau)$ is chosen in such a manner as to yield
a value for $m_b(m_b)$ in its ``observed'' range of
$4.25\pm 0.10\ GeV$ [6], a rather good prediction for the
top-quark mass parameter $m_t(m_t)$ is obtained, which
with the present central value of $\alpha_S(M_Z)=0.12$
lies in the range favoured by the experimental data [7].
Here $\tan\beta$ is found to
saturate what is considered to be a theoretical upper bound on
its value of $m_t/m_b$ and the Yukawa coupling $h$ is found to come
out to be
rather large $O(1-3)$ with a certain insensitivity
to the exact value since it is near a fixed point of its
evolution.
In $SU(5)$ type unification
where $\tan\beta$ is free, the region $\tan\beta \approx 1$
is also a region which is favoured for the unification of
the b-quark and $\tau$-lepton masses from the observed data [4].
One crucial difference between the two extremes discussed
above is that in the $SO(10)$ case the Yukawa couplings of
the b-quark (and that of the $\tau$ lepton) always remain
comparable to that of the top-quark, with the observed
hierachy in the masses of these quarks arising from the
large value of $\tan\beta$, while in the $SU(5)$ case
the Yukawa couplings of the b-quark and the $\tau$-lepton
are negligible in comparison with that of the top-quark.
Furthermore, with large $\tan\beta$ in the MSSM, the mass of
the lightest higgs boson is expected to be no larger than
$140$ GeV [8].

The above discussion about unification does not involve in any great detail
the remaining aspects of the embedding of the standard model
into a supersymmetric grand unified framework.  The minimal
supersymmetric extension
of the standard model requires, besides the superpartners, the introduction
of an additional Higgs
doublet,
and indeed with this matter content and an additional
symmetry known as matter parity [2], to forbid couplings
that lead to rapid nucleon decay, it is possible to construct
a self-consistent and highly successful
framework of the MSSM.
Despite its many successes it may be premature
to confine our attention only to the MSSM, especially because of the
presence of the dimensionful Higgs bilinear parameter $\mu$ in the
superpotential.
Recently, a systematic study of the simplest alternative to
the MSSM, the NMSSM, in the limit of large $\tan\beta$ was
undertaken [9], wherein the bilinear term  $\mu H_1 H_2$ of the MSSM
is replaced by
\begin{equation}
\lambda S H_1.H_2+\frac{1}{3}k S^3,
\end{equation}
 with the effective
``$\mu$'' term generated by the vacuum expectation value
$<S>(\equiv s)\neq 0$.  This model is particularly interesting since
it does not affect the positive features of the MSSM including
gauge coupling unification [10], and allows a
test of the stability of the features of the MSSM such
as the upper bound on the mass of the lightest Higgs boson
with favourable results.  In Ref. [9] it was shown that
the lightest Higgs boson mass in NMSSM, in this
limit, is $\stackrel{_<}{_\sim}140$ GeV.

In this paper we study the particle spectrum of the NMSSM in the
limit of large $\tan\beta$ and compare it with the corresponding
spectrum obtained in the MSSM.
We carry out a renormalization group analysis of
this model with universal boundary conditions and analyze the
renormalization group improved tree-level potential at the
scale $Q_0$. The cut off scale for the renormalization group evolution
is chosen to be the geometric mean of the scalar top quark masses which
is roughly equal to that of the geometric mean of the scalar b-quark
masses as well, since during the course of their evolution the Yukawa
couplings of the t and b-quarks are equal upto their hypercharges and the
relatively minor contribution of the $\tau$-lepton.
Whereas in the MSSM the parameters $\mu$ and $B$
(the soft susy parameter characterizing the bilinear term in
the scalar potential) do not enter into the evolution of the
other  parameters of the model at one-loop level, the situation
encountered here is drastically different with a systematic
search in the parameter space having to be performed with
all parameters coupled from the outset.
Our analysis of the minimization conditions
that ensure a vacuum gives rise to severe fine tuning problems,
that are worse in NMSSM as compared to the ones that arise in MSSM.
The problems are further compounded by having to satisfy the constraints of
three minimization conditions, rather than two such conditions
that occur in MSSM.  In previous studies of the model where $\tan\beta$ was
free,  the tuning of parameters
was possible in order to meet all the requisite criteria, viz.,
minimization conditions, requirement that the vacuum
preserve electric charge and colour, etc.  However, in
the present case where $\tan\beta$ is fixed and large, what
we find is
a highly correlated system.
\bigskip

\noindent{\bf 2. The Model}

\bigskip

We recall the basic features of the NMSSM in what follows.
The model is characterised by the following couplings in the
superpotential
\begin{equation}
W=h_t Q\cdot H_2 t^c_R+h_b Q\cdot H_1b^c_R+h_\tau L\cdot H_1\tau^c_R+
\lambda S H_1.H_2+\frac{1}{3}kS^3,
\end{equation}
where we have written only the interactions of the heaviest
generation and the Higgs sector (doublet and singlet) of the theory.
In addition, one has to add to the potential obtained from (2) the most
general terms
that break supersymmetry softly which, in the conventions of Ref. [9], are:
$$
(h_tA_t \tilde{Q}\cdot H_2 \tilde{t^c_R}+
h_b \tilde{Q}. H_1 \tilde{b^c_R} +
h_\tau A_\tau \tilde{L} \cdot H_1\tilde{\tau^c_R} +
\lambda A_\lambda H_1\cdot H_2 S +
{{1}\over{3}}k A_k S^3)+ {\rm h. c.}
$$
$$
+m_{H_1}^2|H_1|^2+m_{H_2}^2|H_2|^2+m_S^2|S|^2
+m_{\tilde{Q}}^2|{\tilde{Q}}|^2+
m_{\tilde{t}}^2|{\tilde{t}}^c_R|^2
+m_{\tilde{b}}^2|{\tilde{b}}^c_R|^2
+m_{\tilde{\tau}}^2|{\tilde{\tau}}^c_R|^2.
$$

The conventions for the gaugino masses follow those of the
MSSM [11].  The minimization conditions
(evaluated at $Q_0$ after all the parameters are evolved
via their one-loop renormalization group equations
down to this scale) are:
\begin{eqnarray}
m_{H_1}^2=-\lambda{{v_2}\over{v_1}} s (A_\lambda+ks) -\lambda^2(v_2^2+s^2)
+{{1}\over{4}}(g^2+g'^2)(v_2^2-v_1^2), \\
m_{H_2}^2=-\lambda {{v_1}\over{v_2}} s (A_\lambda+ks)-\lambda^2(v_1^2+s^2)
+{{1}\over{4}}(g^2+g'^2)(v_1^2-v_2^2),\\
m_{S}^2=-\lambda^2(v_1^2+v_2^2)-2k^2s^2-2\lambda s v_1 v_2 - kA_k s -
{{\lambda A_\lambda v_1 v_2}\over {s}}.
\end{eqnarray}
One may rewrite the first two minimization equations to obtain [9]
\begin{eqnarray}
\tan^2\beta={{m_Z^2/2+m_{H_1}^2+\lambda^2s^2}\over
{m_Z^2/2+m_{H_2}^2+\lambda^2s^2}},  \\
\sin 2\beta={{(-2\lambda s)(A_\lambda+k s)}\over{m_{H_1}^2
+m_{H_2}^2+\lambda^2(2s^2+v^2)}}.
\end{eqnarray}
Note that eq. (6) guarantees that, as in the MSSM, $\tan\beta$
must lie between 1 and $m_t/m_b$ [9].
Eq.(6) also shows that, to achieve a large $\tan\beta$, with the
essential degeneracy of $m_{H_1}^2$ and $m_{H_2}^2$ enforced
by the renormalization group equations, the denominator of
the equation has to come out to be small at low scale, implying
 the fine-tuning condition
\begin{equation}
m_{H_2}^2+\lambda^2s^2 \approx -m_Z^2/2.
\end{equation}

\noindent Here one notes that the correspondence with the MSSM will occur
in a certain well defined manner with the identification of
$\lambda s$ with $\mu$.  Similarly one has to identify
$A_\lambda+ks$ with $B$.  It has been shown [9] that in
the large $\tan\beta$ case this identification occurs in a
novel way that is not generic to the model, say, in the limit
of $\tan\beta \approx 1$.
Furthermore, eq. (7) implies
\begin{equation}
A_\lambda\approx -ks.
\end{equation}
This is similar to the condition in the MSSM that $B\approx
0$.  Here the situation is far worse since $A_\lambda$ is not
a parameter that is fixed at $Q_0$ but is present from
the outset.  This is the first of the fine tuning problems
that we encounter.

A rearrangement of eq. (7) yields
\begin{equation}
\lambda s (A_\lambda+ks)=\tan\beta
(-m_{H_2}^2-\lambda^2s^2){{m_Z^2}\over{2}}{{(\tan\beta^2-1)}
\over{(\tan\beta^2+1)}}-{{\lambda^2v^2\sin 2\beta}
\over{2}}.
\end{equation}
In  eq. (10) it is legitimate to discard the last term
for the case of large $\tan\beta$, and one sees here that with the
 identification
of the appropriate parameters in terms of the MSSM parameters as
described earlier, one recovers all the analogous MSSM relations
for all values of the other parameters without having to go
through a limiting procedure[12], as is the case when $\tan \beta$
is arbitrary.

The next fine tuning condition we encounter is related to
the third minimization condition which we rewrite as:
\begin{equation}
m_S^2=-\lambda^2 v^2-2 A_\lambda^2+{{\lambda v^2 A_\lambda\sin 2\beta}
\over{k}}+A_\lambda A_k +{{\lambda k v^2 \sin 2\beta}\over{2}}.
\end{equation}
In order to satisfy eq. (11), viz., that $m_S^2$ come out positive
(see e.g., Fig. 1)
in order to have a physically acceptable ground state,
one must have at least have
large cancellations between the fourth and the
first two terms since the terms proportional to $\sin 2\beta$
are negligible.  This requires that $A_k$ and $A_\lambda$
come out with the same sign and that the product be sufficiently
large.  It has been shown that this condition leads to
problems with finding solutions with sufficiently small trilinear
couplings in magnitude [9].

The first step in the study [9]
is to estimate the scale
$M_X$ with the choice of the SUSY breaking scale $Q_0\sim 1\ TeV$.
For $\alpha_S(m_Z)=0.12$, $Q_0 = 1\ TeV$ and $\alpha=1/128$,
we find upon integrating the one-loop beta functions, $M_X=1.9\times
10^{16}\ GeV$ and the unified gauge coupling $\alpha_G(M_X)=1/25.6$.
Next we choose a value for the unified Yukawa coupling $h$ of $O(1)$.
The free parameters of the model are $(M_{1/2},\ m_0,\ A,\ \lambda,\ k)$,
which are the common gaugino mass, the common scalar mass, the
common trilinear scalar coupling and the two additional Yukawa couplings,
respectively.
Note that our convention [9] requires us to choose $\lambda>0$ and
$k<0$ in order to conserve CP in the Yukawa sector of the model[12].

Writing down the coupled system of renormalization group (RG)
equations for the 24 parameters of the model
[13,14] that are coupled to each other,
including the contributions of $h_b$ and $h_\tau$ [9], we compare the
numerical values of the mass parameters that enter the left hand sides
of the minimization equations (3)-(5) with the combinations of the
parameters that enter the right hand side of these equations as obtained
from the RG evolution.  It turns out that the first of the minimization
conditions eq. (3) is the one which is most sensitive to the choice of
intial conditions
reflecting the fine tuning discussed earlier.  Furthermore, in order
to guarantee the absence of electric-charge breaking vacua, we impose
the constraint $|A|<3 m_0$ [15].  For the NMSSM at large $\tan\beta$
this choice may have to be strengthened further due to the presence
of large yukawa couplings for the b-quark.  The situation is considerably
less restrictive when mild non-universality is allowed and, for
instance, if strick Yukawa unification is relaxed.  Given these
uncertainties, we choose to work with the present constraint [9].

The parameter space of the model is scanned by taking values of the
input parameters $(M_{1/2}, m_0, A, h, \lambda, k)$ at the GUT scale
$M_X$ which  are then evolved down to present energies $Q_0$
through the RG evolution to obtain the values of the crucial
parameters $\tan\beta, r(\equiv s/v), A_\lambda, A_k, k s$ in
addition to the soft masses that appear on the LHS of eq. (3)-(5).
The parameters in Ref. [9] were chosen so as to study,
$r_1,\ r_2$ and $r_3$, the differences
in the LHS and RHS of eqs. (3)-(5), divided by the right-hand
side of each, and to minimize the magnitude of each of these
and then study the change in sign that these suffer as the parameters
are varied.  In Ref. [9]  the enormous difficulties
faced in trying to achieve a simultaneous solution to $r_i=0,\ i=1,2,3$
were noted.  In particular, it was shown that $r_3=0$ required the
presence of values for $|A|/m_0$ of almost 3 or more.  This requirement
will have a profound impact on the particle spectrum under discussion
as we will show below.

We also ensure that the value of $\Delta E$, the difference in the
value of the scalar potential computed with the scalar fields attaining
their vacuum expectation values, $(v_1, v_2, s)$ and its value computed
at the symmetric minimum is negative making the $SU(2)\times U(1)$
breaking minimum energetically favourable.  Furthermore, the squared
mass of the charged Higgs boson [12] was also computed in order to single out
only those points in the parameter space where it is positive in order
to not break electric charge spontaneously.  The intricate relationship
between the various free parameters enforced by the rough scale invariance
enjoyed by the evolution equations has been emphasized [9].  Finally,
we note that $r$, the ratio of the singlet to the doublet
expectation value persistently remains large for the choice of parameters
considered with large $\tan\beta$, corresponding to vacuum expectation value
of the singlet $s$ being of the order of $10$ TeV.  This is
substantially different
from what happens in the case of $\tan\beta \simeq 1$ [12].
We note that the singlet
expectation value is not constrained by the experimental data.

\bigskip

\noindent{\bf 3. Results and Discussion}

\bigskip

Having described the NMSSM with large $\tan\beta$ in some detail
in the previous section [9], we now turn to obtaining
the particle spectrum of the model.

In Fig. 1, we show a typical evolution of the three soft SUSY
breaking mass parameters $m_{H_1}^2$,  $m_{H_2}^2$ and  $m_{S}^2$
from $M_X$ down to the low scale $Q_0$ with a choice of parameters
such that all constraints are satisfied and we are in the neighbourhood
of an $SU(2)\times U(1)$ breaking vacuum.  We note that because of
the possibility of large value of $m_t(m_t)=181$ GeV, we have
a large value for $h$ so that the Yukawa couplings dominate the
evolution of these parameters over the gauge couplings.  This in turn
forces the mass parameters to remain large at large momentum scales
compared to their values at smaller momentum scales.

In supersymmetric theories with R-parity conservation, the lightest
supersymmetric particle is generally assumed to be the neutralino.
In NMSSM, the neutralino mass matrix is a $5\times 5$ matrix
whose general properties are  discussed in [16].  The parameters
that determine the mass matrix are $\lambda, k, s, \tan\beta, M_1$
and $M_2$, where $M_1$ and $M_2$ are the SUSY breaking bino and
wino masses, respectively.  Choosing the input parameters at the
GUT scale so that all the constraints are satisfied in the manner
detailed in the previous section, we obtain the values
of the parameters which enter the neutralino mass matrix at the
weak scale.  We then evaluate the neutralino mass matrix numerically.
The chargino mass matrix, on the other hand, is the same as in the MSSM with
$\mu$ replaced by $\lambda s$.  The chargino masses can, therefore,
be obtained analytically.

One result of this procedure is shown in Fig. 2, where we plot the
lightest neutralino masses for a specific choice of input parameters
(see table 1), which are similar to ones detailed in Ref. [9],
as a function of the top quark mass.
{}From our scan of the parameter space, we have found
that the lightest neutralino is
almost a pure bino in the limit of large $\tan\beta$.  Furthermore,
all other neutralinos except the heaviest one have a negligible
singlet component.  This indicates that the singlet completely
decouples from the lighter neutralino spectrum.  The mass of the
lightest  neutralino in this case
is determined by the simple mass relation for the bino
\begin{equation}
M_1\simeq \left( {\alpha_1(M_G)\over \alpha_G}\right) M_{1/2}
\simeq 0.45 M_{1/2}.
\end{equation}
The masses of the heavier neutralinos lie in the range of $0.5-1$ TeV.
Following the same procedure, as in the
case of the neutralinos, of obtaining the parameters entering
the chargino mass matrix from the RG evolution of parameters at the
GUT scale, we obtain the lightest chargino mass as a function of
$m_t(m_t)$ also displayed in Fig. 2.  The lightest chargino mass
bears a relation to $M_{1/2}$ similar to the neutralino relation,
with $\alpha_1$ in eq. (12) replaced by $\alpha_2$ reflecting that
it is primarily a charged wino.  The heavier chargino mass is found to
be $\simeq 1$ TeV.  The gluino mass is found to be $1.6$ TeV for
the choice of parameters of Fig. 2 and follows from a relation
similar to eq. (12) with  $\alpha_1$ replaced by  $\alpha_S$.

We come now to the spectrum of CP-even Higgs bosons of the model.
The mass matrix of these Higgs bosons in a $3\times 3$ matrix
which has been discussed extensively in the literature [17,18,19].
Nevertheless, in order to understand the quantitative features
of the results we have obtained for the lightest CP-even higgs
we need to go into some detail regarding the actual choices of parameters
entering the computation and the correlations between the various elements
of the spectrum.  Such a discussion has been made available [8]
for the case of the MSSM and we will offer a comparison for the
model at hand in the present case.
In Fig. 3, we plot for typical and reasonable values of the input
parameters in the region where the vacuum is expected to lie, the
mass of the lightest CP-even Higgs boson as a function of the
top-quark mass $m_t(m_t)$ in the range that is most favoured under
these boundary conditions [3,8,20,21].
The choice of parameters here is closely related to the family
of solutions studied extensively in Ref. [9] and would serve
as a typical example of the numbers we have explored.  Also,
the features seen in
this figure may be understood in terms of several of the entries
appearing in Table 1 and in particular with those of the corresponding
entries for the CP-odd neutral higgs masses.  In the MSSM, for
instance, it is well known that when the mass of its unique CP-odd
boson $m_A >> m_Z$, the substantive part of the radiative correction
is picked up by the lighter of the CP-even bosons, $h^0$.
As $m_A$ approaches
$m_Z$, the radiative corrections are now shifted to the heavier
of the CP-even higgs bosons, $H^0$.  Such a feature is observed here:
for those choices of parameters in Table 1 that yield a somewhat smaller
$m_{P_1}$, we find that the radiative corrections to the lighest
CP-even higgs, $h_1$ are smaller.  Due to the complexity of the system
under investigation and the difficulty to control precisely the
numerical confidence in the choice of the parameter $\lambda$ for
a given $h$ with all other parameters held fixed, we do not know
how precisely close the choice of parameters of Table 1 are to
a genuine ground state.
Furthermore, we note that
the clarity with which the correlations have been
observed in the MSSM between $m_A$ and $m_{h^0}$ does not have
a simple parallel here due to the presence of a large number of
physical states.
A more precise, albeit prohibitively time consuming, determination could then
ensure that the spurious wobble seen in Fig. 3 is eliminated
and establish a more reliable correlation between increasing
$h$ and the rise of the mass of $h_1$ and the correlations
with $m_{P_1}$.
Furthermore, a refinement of the choice of parameters when
a more complete computation based on the minimization of the
one-loop effective potential could stablize the figures presented
here.

We note that the lightest
Higgs bosons mass $\sim 130$ GeV for a wide range of parameters
which nearly saturates the upper bound of $140$ GeV [9].  The mass of the
lightest Higgs boson in the NMSSM lies in the same range as in
the MSSM with large $\tan\beta$.  This is a consequence of the
the largeness of $\tan\beta$:  the contribution to the tree level
mass which depends on the tri-linear coulings $\lambda$ is small,
being proportional to $\sin^2 2\beta$, so that the upper bound on
the lightest Higgs mass reduces to the corresponding upper bound
in the MSSM when appropriate identification of parameters is made.
We also note that the upper bound on the lightest Higgs mass depends
only logarithmically on $r$, and hence on the singlet vacuum
expectation value $s$, in the limit of large $r$, which, therefore
decouples from the bound [22].  Furthermore, the lightest Higgs
boson in almost a pure doublet Higgs (${\rm Re}\ H_2^0$), with
the singlet component being less that $1\%$ in the entire range
of parameters considered.  It is only the second heavier CP  even
Higgs boson $h_2$ that is predominantly a singlet.  Its mass ranges
between $740$ GeV and $2.3$ TeV.  The heaviest
CP even Higgs boson $h_3$ is again
predominantly a doublet Higgs  (${\rm Re}\ H_1^0$) with its mass
varying between $4-6$ TeV.  This implies that all the CP-even
Higgs bosons, except the lightest one, decouple from the spectrum.
The results presented above, that the lightest Higgs boson is
almost purely a doublet Higgs at large $\tan\beta$ is in contrast
to the situation with low values of $\tan\beta$, where the
lightest CP-even Higgs boson contains a large admixture of the
gauge singlet field $S$ [12,23,24].

On the other hand, we note from Table 1 that the two
CP-odd Higgs bosons,
$P_1$ and $P_2$ in the model are heavy, their masses being in the range
$2$ TeV and $6$ TeV, respectively.  Also, the lightest CP-odd
state is predominatly a Higgs singlet, thereby effectively
decoupling from the rest of the spectrum.  The charged Higgs boson
mass $m_C$ lies, for most of the cases that we have studied, in the range
$1-2$ TeV.

In order to discuss the feature of the remainder of the spectrum,
viz., the sfermion spectrum let us first recall some features of
the spectrum of the MSSM.
In MSSM, it has been observed [8] that the presence of large
Yukawa couplings for the b-quark as well as the $\tau$ lepton
and also the presence of large trilinear couplings
could lead to the lighter of the scalar $\tau$'s tending to become
lighter than the lightest neutralino, which is the most favored
candidate in such models for the lightest supersymmetric particle.
In particular, in order to overcome such cosmological constraint
for given values of $M_{1/2}$, lower bounds on $m_0$ were found to
emerge.  In turn, increasing $m_0$ implies ever decreasing $m_A$
thus leading to further constraints on the parameter space of the MSSM [8].
The competing tendencies between the lighter scalar tau mass and $m_A$ have
been shown to play an extraordinary role in MSSM for large values of
$\tan\beta$ in establishing a lower bound $\sim 450$ GeV on $M_{1/2}$. Given
the
complexity of the system of equations, it has not been possible to
to extract similar lower bounds on $M_{1/2}$ in NMSSM. Nevertheless, in Ref.
[9]
the intimate link between the ground states of the two models has been
established and a much more sophisticated and time consuming analysis of
the present model is also likely to yield a lower bound that is unlikely
to be very different from the one obtained in MSSM. As a result, in confining
ourselves to numbers of this magnitude and higher, we find a heavy spectrum.
More recently [25] further experimental constraints on MSSM have
been taken into account resulting in an extension of the minimal assumptions
at $M_X$ by including non-universality for scalar masses.
Indeed, in the present analysis similar problems have been encountered
with some of the choice of parameters studied in Ref. [9], with
$m_{\tilde{\tau}_1}$ tending to lie below the mass of the lightest
neutralino due to the persistent presence of large Yukawa couplings
and more so due to that of the large trilinear couplings.
Nevertheless, given the fact that the present work minimizes the
tree-level potential, and that the violations of cosmological
constraints are not serious in that minor adjustments of
$|A/m_0|$ solve this problem efficiently,
we consider the regions of the
the parameter space we have studied
to be reasonable ones.  Furthermore,
it could be that the extension of minimal boundary conditions
along the lines of Ref. [25]
could provide alternative and elegant solutions to
this problem, while preserving the existence
of relatively light scalar $\tau$'s as a prediction of the unification of
Yukawa couplings in both the minimal and nonminimal
supersymmetric standard models.

The heaviest sfermions in the spectrum of NMSSM  as in
MSSM are the scalar quarks.  The scalar quarks here, as in the
MSSM, tend to be much heavier, in the TeV range.  The $SO(10)$ property
that the scalar b-quarks are as massive as the scalar top-quarks is
preserved in the NMSSM.

\bigskip

\noindent{\bf 4. Conclusions}

\bigskip

In the detailed analysis presented here, we have shown that, except
for the lightest CP-even Higgs boson, all the particles implied by
supersymmetry are heavy for large values of $\tan\beta$.  The
gauge singlet field S decouples both from the lightest Higgs
boson as well as the neutralinos.  This is in contrast to the situation
that one obtains for the model at low values of $\tan\beta$.
The LSP of the model continues to be, as in MSSM, the lightest
neutralino that is primarily a bino in composition, with the lighter
scalar $\tau$ with a mass in the neighbourhood of the LSP mass.
The remainder of the spectrum tends to be heavy, from 1 to a few TeV.
We note that the NMSSM in the large $\tan\beta$ regime
rests on a delicately hinged system of
equations and constraints.  Although it provides a good testing
ground for the stability of the predictions of the MSSM, in
practice it deserves great care in its treatment.

\newpage

\noindent{\bf Acknowledgements}:
The research of BA has been supported by the Swiss National
Science Foundation.
PNP thanks the Alexander von Humboldt-Stiftung
and Universit\"at Kaiserslautern, especially
Prof. H. J. W. M\"uller-Kirsten, for support
while this work was completed.
The work of PNP is supported by the
Department of Science and
Technology, India under
Grant No. SP/S2/K-17/94.

\newpage

\noindent{\Large{\bf Table Caption}}

\bigskip

\noindent {\bf Table 1.}  Sample of points in the parameter space
and the computed values of different mass parameters [all masses
in units of GeV].

\bigskip

\noindent{\Large{\bf Figure Captions}}

\bigskip

\noindent {\bf Fig. 1}  The evolution of soft supersymmetry breaking
mass parameters from the grand unified scale $M_X$ to $Q_0$ defined
in the text.  The input parameters are $M_{1/2}=-700$, $m_0=800$ and
$A=1600$ (all in GeV).  The other parameters are $h=1.5,\
\lambda=0.40$ and $k=-0.10$.  The associated value of the top
quark mass is $181$ GeV.

\bigskip

\noindent {\bf Fig. 2} The lightest neutralino and chargino masses
as a function of $m_t(m_t)$.  The input parameters are $M_{1/2}=-700,
\ m_0=800,\ A=2200$ (all in GeV), with the remaining parameters
varied to guarantee a solution.

\bigskip

\noindent {\bf Fig. 3} The lightest CP-even Higgs boson mass as a function
of $m_t$.  The range of parameters is as in Fig. 2.

\bigskip

\bigskip

\newpage

\vskip 5cm

\begin{footnotesize}
$$
\begin{array}
{||c|c|c|c|c|c|c|c|c|c|c|c|c||}\hline
 M_{1/2} & m_0 &  A & h & \lambda & k & m_t(m_t) & m_{h_1}
 & m_{h_2} & m_{h_3} & m_{P_1} & m_{P_2} & m_{C} \\ \hline
-700 & 800 & 2200 & 0.75 & 0.1 & -0.1 & 170.5 & 131.0 & 2314 & 6039 & 3048 &
6918 & 5077 \\
-700 & 800 & 2200 & 1.00 & 0.2 & -0.1 & 176.5 & 132.5 & 1326 & 4822 & 2479 &
6216 & 3031 \\
-700 & 800 & 2200 & 1.25 & 0.3 & -0.1 & 179.6 & 130.3 & 977 & 4342 & 2238 &
5990 & 1862 \\
-700 & 800 & 2200 & 1.50 & 0.4 & -0.1 & 181.4 & 119.0 & 774 & 4297 & 2099 &
6394 & 949 \\
-700 & 800 & 2200 & 1.75 & 0.4 & -0.1 & 182.5 & 133.0 & 1031 & 5116 & 2275 &
6800 & 2901 \\
-700 & 800 & 2200 & 2.00 & 0.5 & -0.1 & 183.3 & 128.0 & 858 & 4823 & 2153 &
6636 & 2227 \\
-700 & 800 & 2200 & 2.25 & 0.6 & -0.1 & 183.8 & 119.0 & 740 & 4644 & 2067 &
6557 & 1668 \\ \hline
\end{array}
$$
\end{footnotesize}
$$
{\rm Table \ 1}
$$
\end{document}